\documentclass[journal]{IEEEtran}
\usepackage{amsmath}
\usepackage{amssymb}
\usepackage{amsfonts}
\usepackage{graphicx}
\usepackage{enumerate,color,graphicx,fancybox,pifont,epsf,epsfig,subfigure,amsmath,amssymb,psfrag}
\usepackage{algorithm}
\usepackage{algorithmic}
\usepackage{cite}
\newcounter{MYtempeqncnt}

\newtheorem{prob}{\textbf{Problem}}

\newcommand{\secref}[1]{Section~\ref{#1}}

\newcommand{\figref}[1]{Figure~\ref{#1}}

\newcommand{\probref}[1]{Problem~\ref{#1}}

\newcommand{\remref}[1]{Remark~\ref{#1}}

\newcommand{\algref}[1]{Algorithm~\ref{#1}}

\newtheorem{rem}{\textbf{Remark}}

\newcommand{\qed}{\nobreak \ifvmode \relax \else
      \ifdim\lastskip<1.5em \hskip-\lastskip
      \hskip1.5em plus0em minus0.5em \fi \nobreak
      \vrule height0.75em width0.5em depth0.25em\fi}

\title{Analysis-by-Synthesis Quantization for \\ Compressed Sensing Measurements}
\author{Amirpasha Shirazinia, \textit{Student Member, IEEE}, Saikat Chatterjee, \textit{Member, IEEE}, Mikael Skoglund, \textit{Senior Member, IEEE} \\}

\begin{document}
\maketitle

\begin{abstract}
  We consider a resource-limited scenario where a sensor that uses
  compressed sensing (CS) collects a low number of measurements in order to
  observe a sparse signal, and the measurements are subsequently
  quantized at a low bit-rate followed by transmission or storage. For such
  a scenario, we design new algorithms for source coding with the
  objective of achieving good reconstruction performance of the sparse
  signal. Our approach is based on an analysis-by-synthesis principle
  at the encoder, consisting of two main steps: (1) the synthesis step
  uses a sparse signal reconstruction technique for measuring the
  direct effect of quantization of CS measurements on the final sparse
  signal reconstruction quality, and (2) the analysis step decides
  appropriate quantized values to maximize the final sparse signal
  reconstruction quality. Through simulations, we compare the
  performance of the proposed quantization algorithms vis-a-vis
  existing quantization schemes.
\end{abstract}

\section{Introduction} \label{sec:intro}

Based on a model of under-determined linear set of equations,
compressed sensing (CS) \cite{08:Candes} aims to reconstruct a
high-dimensional sparse source vector, where most of coefficients are
zero, from an under-sampled low-dimensional measurement vector. With a
limited number of measurements (or a limited sampling resource), CS
has emerged as a new powerful tool for sparse signal acquisition,
compression and reconstruction. For many practical applications, CS
measurements need to be quantized into a finite resolution
representation, for subsequent transmission or storage. Later,
using the finite resolution measurements, sparse signal reconstruction
is performed, potentially followed by other inference tasks.

In this paper, we consider application scenarios where both
measurement (or sampling) and quantization resources are limited. In
particular, we assume that the total number of available bits for
quantization is limited. One scenario would be in wireless body sensor
networks (WBSN) \cite{10:Yang} where a low number of measurements and
a low bit-rate are available due to sensing costs and channel transmission rate
restrictions, respectively.
\textit{Considering availability of limited number of measurements and
  quantization bits, we design new quantization algorithms with the
  objective of achieving high quality sparse signal reconstruction
  from the quantized CS measurements.}

\subsection{Related Works}

CS with quantized measurements has recently started to gain
significant attention in the literature, and the available approaches
are of three main kinds, described below.
\begin{enumerate}
\item In the first category, extensions to existing sparse
  reconstruction schemes have been studied without changing the
  quantization algorithms. Sparse signal reconstruction from noisy
  measurements -- which can be thought of as the effect of
  quantization when the noise is bounded and additive -- has been
  addressed in \cite{06:Candes2}. In \cite{10:Sinan}, CS with finely
  quantized measurements using Sigma-Delta modulation is studied. In
  \cite{10:Zymnis}, the authors focus on convex optimization-based CS
  reconstruction from a set of quantized measurements, whereas
  \cite{11:Dai} considers a greedy search algorithm for this
  purpose. In \cite{11:Jacques}, a reconstruction scheme for more
  precise reconstruction of sparse signals from quantized measurements
  has been developed. In addition, robust schemes (against measurement
  noise) for reconstructing a sparse source from 1-bit quantized
  measurements have been proposed in \cite{12:Yan,12:Jacques}, and the
  design of message-passing algorithms for estimation from quantized
  CS samples has been studied in \cite{12:Kamilov}.
\item The second category considers the trade-off between the aspects
  of quantization (e.g., quantization bit-rate) and CS (e.g., number of
  measurements and loss in sparse reconstruction). In \cite{08:Goyal},
  high bit-rate theoretical bounds on average \textit{end-to-end
    distortion} due to quantization of sparse sources and CS
  measurements have been derived, whereas the goal in \cite{11:Dai} is
  to find high bit-rate average \textit{quantization distortion}
  bounds caused by vector and scalar quantization of CS
  measurements. The aim in \cite{12:Laska} is to analyze a trade-off
  between number of measurements and quantization bit-rate by introducing
  quantization compression regime versus CS compression regime.
\item Finally, in the third category, the main focus is on quantizer
  design for CS measurements while sparse reconstruction schemes are
  fixed. The design of high bit-rate quantizers for CS measurements
  that are optimal with respect to mean square error (MSE) of a
  particular convex optimization-based reconstruction method (LASSO
  \cite{96:Tibshirani}) is studied in \cite{09:Sun}. Moreover, an
  optimal high rate quantizer design under message-passing
  reconstruction algorithms has been presented in
  \cite{11:Kamilov}. In \cite{12:Boufounos}, a framework for scalar
  quantization of CS measurements has been proposed which provides
  exponential decay for instantaneous signal reconstruction
  distortion as a function of bit-rate.
\end{enumerate}

\subsection{Contributions of the Paper}

The main contribution of this paper is in the third category mentioned
above, i.e., quantizer design for CS measurements while sparse
reconstruction algorithms are fixed. For this purpose, choosing the
MSE as a performance criterion, we first derive necessary encoding
conditions so as to minimize reconstruction MSE for a sparse input
vector under a fixed decoder and a sparse reconstruction
algorithm. Then, in order to feasibly implement the resulting rules,
we develop a new framework for scalar quantization of CS
measurements with the objective of achieving a lower \textit{end-to-end
  reconstruction distortion} for the sparse source rather than
\textit{quantization distortion} for CS measurements. Technically,
given a fixed quantizer look-up table and a fixed (but generic) sparse
reconstruction scheme, the proposed algorithms strategically use a
two-step mechanism in a closed-loop fashion: (1) the
\textit{synthesis} step employs a sparse signal reconstruction
technique for measuring the direct effect of quantization of CS
measurements on the final sparse signal reconstruction quality, and
(2) the \textit{analysis} step is performed followed by the
synthesis step in order to choose appropriate quantized values to minimize
the final sparse signal reconstruction distortion. This closed-loop
strategy is known as \textit{analysis-by-synthesis} (AbS) which has
been widely used in multi-media coding
\cite{88:Kroon,95:Aizawa,97:George}.

To the best of our knowledge, the AbS approach has not been used for
quantization of CS measurements earlier. The use of AbS principle is
shown to provide a significantly better reconstruction performance
compared to schemes that only consider direct quantization of CS
measurements. Here, we mention that AbS requires higher computation. We analyze the computational complexity of the proposed algorithms, where it is
shown that the complexity depends on availability of two compression
resources, i.e., quantization bit-rate and number of CS
measurements. As a byproduct, we also propose a low complexity
scheme based on quantization of estimated sparsity patterns at the
quantizer encoder which performs well at high quantization bit-rates. Further, we develop an \textit{adaptive} quantization method by combining the proposed schemes in order to
provide high-quality performance at all ranges of quantization and
measurement rates. We experimentally evaluate the performance of our proposed algorithms,
and also compare them with that of existing schemes for quantization of CS
measurements.

\subsection{Outline of the Paper}

The remaining parts of the paper are organized as follows. In
\secref{sec:problem}, we give the problem statement which involves the
CS and quantization models and also performance criterion. In
\secref{sec:quant}, we propose the new algorithms, and analyze their computational complexities. Thereafter, we show an adaptive quantization scheme in \secref{subsec:adaptive coding}. The numerical results are given in
\secref{sec:numerical}, and the conclusions are drawn in
\secref{sec:conclusion}.

\subsection{Notations} \label{subsec:notations} Scalar random
variables (RV's) will be denoted by upper-case letters while their
realizations (instants) will be denoted by the respective lower-case
letters. Random vectors of dimension $n$ will be represented by
boldface characters. Hence, if $\mathbf{Z}$ denotes a random row
vector $[Z_1,\ldots,Z_n]$, then, $\mathbf{z} = [z_1,\ldots,z_n]$
indicates a specific realization of $\mathbf{Z}$. We will denote a
sequence of RV's $J_1,\ldots,J_N$ by $\mathbf{J}_1^N$, further,
$\mathbf{J}_1^N = \mathbf{j}_1^N$ implies that
$J_1=j_1,\ldots,J_N=j_N$. Matrices will be denoted by capital Greek
letters, and their pseudo-inverse by $(\cdot)^\dag$. Further, a set is
shown by a calligraphic character and its cardinality by $|\cdot|$,
for example, $\mathcal{A} = \{a_i\}_{i=1}^n$ represents a set with
cardinality $|\mathcal{A}|=n$. We will also denote the transpose of a
vector by $(\cdot)^T$.  We will use~$\mathbb{E}[\cdot]$ to denote the
expectation operator. The $\ell_p$-norm ($p > 0$) of a vector
$\mathbf{z}$ will be denoted by $\|\mathbf{z}\|_p = (\sum_{n=1}^N
|z_n|^p)^{1/p}$. $\|\mathbf{z}\|_0$ represents $\ell_0$-norm which is
the number of non-zero coefficients in $\mathbf{z}$. Finally, for two real
functions $f(n)$ and $g(n)$, $f(n) = \mathcal{O}(g(n))$ if $\exists c
\in \mathbb{R}^+$ and $0 \leq f(n) \leq c g(n)$.

\section{Problem Statement} \label{sec:problem}

In this section, we first introduce the preliminaries of CS and
quantization, and then the design criterion and goal will be
described.

\subsection{Preliminaries on the CS Framework} \label{subsec:formulation}

The CS context can be thought of as a source compression method where
a high-dimensional vector is mapped to a lower-dimensional vector both
belonging to arbitrary continuous sets. Formally, we let a random sparse
(in a fixed basis) signal $\mathbf{X} \in \mathbb{R}^M$ be linearly
encoded using a known sensing matrix $\mathbf{\Phi} \in \mathbb{R}^{N
  \times M}$ ($N < M$) representing a measurement (sampling) system
which results in an under-determined set of linear measurements, i.e.,
\begin{equation}
    \mathbf{Y = \Phi X} \in \mathbb{R}^N.
\end{equation}
We let $\mathbf{X}$ be a $K$-sparse vector, i.e., it has at most $K$ ($K < N$) non-zero coefficients, where the location of the non-zero's are uniformly drawn from all ${M \choose K}$ possibilities, and the magnitude of the non-zero coefficients are identically and independently drawn from a
known distribution. 
We define the support set of the sparse vector $\mathbf{X}
= [X_1,\ldots,X_M]^T$ by $\mathcal{S} \triangleq \{m : X_m \neq 0 \}
\subset \{1,\ldots,M\}$ and $|\mathcal{S}| = \|\mathbf{X}\|_0 \leq K$.

For the purpose of estimating a sparse vector from (possibly noisy) measurements, several efficient techniques have been developed based on convex
optimization methods (see e.g.~\cite{98:Chen,06:Candes2,07:Candes}),
iterative greedy search algorithms (see
e.g. \cite{07:Tropp,08:Blumensath,09:Dai,09:Needell,12:Saikat}) and
Bayesian estimation approaches (see
e.g. \cite{07:Larsson,08:Ji,09:Elad,10:Protter,12:Kun}). For example,
the goal of iterative greedy search algorithms is to first detect the
sparsity pattern, i.e. location of non-zero coefficients, and then
estimate the unknown non-zero coefficients. \textit{In this paper, the design
we present is generic in the sense that it works for any (fixed)
sparse reconstruction algorithm.} We denote a sparse reconstruction
algorithm by a mapping function $\textsf{R}:\mathbb{R}^N \rightarrow \mathbb{R}^M$ which takes a measurement vector in $N$-dimensional space as an input, and produces an estimate of the sparse source vector in $M$-dimensional space ($N<M$) through a highly non-linear procedure. We also assume that the sensing matrix $\mathbf{\Phi}$ is provided for the sparse reconstruction.

\subsection{Quantization of CS Measurements} \label{subsec:SQ}

Scalar quantization is the process of mapping a sample from a
continuous set to a discrete alphabet. We consider scalar quantization
of the random CS measurements $Y_n$'s ($n=1,\ldots,N$). For this
purpose, quantization is divided into \textit{encoding} and
\textit{decoding} tasks. We consider a scalar \textit{quantizer
  encoder} which maps each measurement to an appropriate index in a
finite integer set in order for a \textit{quantizer decoder} to make
an estimate of the measurements based on the received index and a
known decoding look-up table. We assume that the total bit budget
(rate) allocated for quantization is $R_x \triangleq M r_x$ bits per
vector $\mathbf{X}$ in which $r_x \in \mathbb{R}^+$ is the assigned
quantization bit-rate to a scalar component of $\mathbf{X}$. Having the
observations $\mathbf{Y = \Phi X}$, each scalar $Y_n$ ($n=1,\ldots,N$) of the measurement vector $\mathbf{Y}$ is encoded by $r_y \triangleq Mr_x/N$
bits.\footnote[1]{In practice, $r_y$ can be a non-integer value,
however, in the design procedure we let $r_y$ be a positive integer,
and later in simulation results, we show how to address the
non-integer issue.} For each entry $Y_n$, the quantizer encoder is
defined by a mapping $\textsf{E}: \mathbb{R} \rightarrow \mathcal{I}$,
where $\mathcal{I}$ denotes the index set defined as $\mathcal{I}
\triangleq \{0,1,\ldots,2^{r_y}-1\}$ with $|\mathcal{I}| =
2^{r_y}$. Denoting the quantized index by the RV $I_n$ ($n
=1,\ldots,N$), the encoder works according to $Y_n \in
\mathcal{R}^{i_n} \Rightarrow I_n = i_n$, where the sets
$\{\mathcal{R}^{i_n}\}_{i_n=0}^{2^{r_y}-1}$ are called \emph{encoder regions}
and $\bigcup_{i_n=0}^{2^{r_y}-1} \mathcal{R}^{i_n} = \mathbb{R}$. Next, we define the quantizer decoder which is characterized by a mapping $\textsf{D}:
\mathcal{I} \rightarrow \mathcal{C}_n$. The quantizer decoder takes
the index $I_n$, and performs according to an available look-up table;
$I_n = i_n \Rightarrow \widehat{Y}_n = c_{i_n}$. Note that $\widehat{Y}_n$ is the quantized measurement RV
associated with the entry $Y_n$, and the set of all reproduction
\textit{codepoints} $\mathcal{C}_n \triangleq
\{c_{i_n}\}_{i_n=0}^{2^{r_y}-1}$ associated with the entry $Y_n$ is
called a \textit{codebook}. Let us denote by $\widehat{\mathbf{X}} \in
\mathbb{R}^M$ the estimation of the input signal vector from quantized
measurement vector $\widehat{\mathbf{Y}} \triangleq
[\widehat{Y}_1,\ldots,\widehat{Y}_N]^T$ using a sparse reconstruction
function \textsf{R}. Then in a more compact way, given a fixed reconstruction $\textsf{R}$, we state acquisition,
quantized transmission and reconstruction equations as
\begin{equation} \label{eq:equations}
\begin{aligned}
    &\hspace{2.2cm} \mathbf{Y} = \mathbf{\Phi X},& \\
    &\hspace{0.6cm} I_n = \textsf{E}\left(Y_n\right) , \hspace{0.1cm}\widehat{Y}_n =  \textsf{D}\left(I_n\right)=c_{I_n}, \forall n & \\
    &\hspace{0cm} \widehat{\mathbf{X}}(I_1,\ldots,I_N) \triangleq \widehat{\mathbf{X}}(\mathbf{I}_1^N) = \textsf{R}\left([c_{I_1},\ldots,c_{I_N}]^T\right).&
\end{aligned}
\end{equation}

\subsection{Objectives and Preliminary Analysis} \label{subsec:criteria}

We establish the goal of our work as follows.
\begin{prob} \label{prob1} For the system setup \eqref{eq:equations} with  fixed codebook sets $\mathcal{C}_n =
  \{c_{i_n}\}_{i_n=0}^{2^{r_y}-1}$ ($n=1,\ldots,N$), a bit budget
  $R_x = Mr_x = N r_y$, the objective is to implement the encoder, i.e., to find \textit{encoding indexes} (transmission indexes) $i_n \in
  \mathcal{I}$ ($n=1,\ldots,N$) such that the \textit{end-to-end} MSE,
  $\mathbb{E}[\|\mathbf{X - \widehat{X}}\|_2^2]$, is
  minimum. Therefore, we address the
  following optimization problem
\begin{equation} \label{eq:opt e2e}
    \{i_1^\star,\ldots,i_N^\star\} = \underset{\{i_n \in \mathcal{I}\}_{n=1}^N}{\textrm{arg min }} \mathbb{E} [\|\mathbf{X} - \widehat{\mathbf{X}} \|_2^2 ],
\end{equation}
where $\{i_n^\star\}_{n=1}^N$ are the optimal encoding indexes (with respect to minimizing the end-to-end MSE) for quantization of the measurement vector $\mathbf{Y}=[Y_1,\ldots,Y_N]^T$.
\end{prob}

To clarify the objective, we first introduce a common
encoding approach, referred to as \textit{nearest-neighbor coding} for CS measurements. In this type of quantization, each scalar entry of the measurement vector is coded to its nearest
codepoint. Therefore, given a fixed codebook associated with the
measurement entry $Y_n=y_n$, and given that $\widehat{Y}_n = c_{i_n}$,
the nearest-neighbor quantizer/encoder uses the following encoding
rule
\begin{equation} \label{eq:nearest n}
    i_n^\ast = \underset{i_n \in \mathcal{I}}{\textrm{arg min }} |y_n - c_{i_n}|^2, \forall n ,
\end{equation}
which minimizes the MSE per measurement entry, i.e., $\mathbb{E}[|Y_n - \widehat{Y}_n|^2]$. However, this approach does not necessarily guarantee that the
end-to-end MSE (the final performance measure) $\mathbb{E}
[\|\mathbf{X} - \widehat{\mathbf{X}} \|_2^2]$ is also minimized
subject to fixed codebook sets. This is due to non-linear behavior of
the sparse reconstruction function $\textsf{R}$ and non-orthogonality of the CS
sensing matrix $\mathbf{\Phi}$. We also mention that, in this work, the
decoder codebooks $\mathcal{C}_n$ ($n=1,\ldots,N$) are designed off-line, and we do not address the
separate issue of codebook design; the codebooks are given and
fixed.

Now we show how MSE-minimizing encoding indexes are
chosen. First, let us denote the \textit{minimum mean square error}
(MMSE) estimation of $\mathbf{X}$ given the measurements
$\mathbf{Y} = \mathbf{y}$ by
\begin{equation} \label{eq:MMSE est}
    \widetilde{\mathbf{x}}(\mathbf{y}) \triangleq \mathbb{E}[\mathbf{X|Y=y}] \in \mathbb{R}^M.
\end{equation}
Next, using the notations introduced in \secref{subsec:notations}, we
rewrite the end-to-end MSE, $\mathbb{E}[\|\mathbf{X} -
\widehat{\mathbf{X}}\|_2^2]$, as \eqref{eq:suboptimal condition} on top of next page,
where $(a)$ and $(b)$ are followed by marginalization over $\mathbf{Y}$ and $\mathbf{I}_1^N$, respectively. Also, $(c)$ follows from interchanging the integral and summation and the fact that $\textrm{Pr} \{\mathbf{I}_1^N=\mathbf{i}_1^N |\mathbf{Y=y}\} = 1$, $\forall y_1 \in \mathcal{R}^{i_1}, \ldots, y_N \in \mathcal{R}^{i_N}$, and otherwise the probability is zero. Here, $f(\mathbf{y})$ is the $N$-fold probability density function (p.d.f.) of the measurement vector.

\begin{figure*}[!t]
\normalsize
\setcounter{MYtempeqncnt}{\value{equation}}
\begin{equation} \label{eq:suboptimal condition}
\begin{aligned}
  \mathbb{E}[\|\mathbf{X} - \widehat{\mathbf{X}} \|_2^2] &= \mathbb{E}[\|\mathbf{X} - \widehat{\mathbf{X}}(\mathbf{I}_1^N) \|_2^2] \stackrel{(a)}{=} \int_{\mathbf{y}} \mathbb{E}[\|\mathbf{X} \!-\! \widehat{\mathbf{X}}(\mathbf{I}_1^N) \|_2^2 | \mathbf{Y=y}] f(\mathbf{y}) d \mathbf{y}  & \\
  &\stackrel{(b)}{=} \int_{\mathbf{y}} \sum_{i_1} \ldots \sum_{i_N} \textrm{Pr}\{\mathbf{I}_1^N = \mathbf{i}_1^N | \mathbf{Y=y}\} \mathbb{E}[\|\mathbf{X} - \widehat{\mathbf{X}}(\mathbf{I}_1^N) \|_2^2 | \mathbf{Y=y},\mathbf{I}_1^N = \mathbf{i}_1^N] f(\mathbf{y}) d \mathbf{y}& \\
  &\stackrel{(c)}{=}  \sum_{i_1} \ldots \sum_{i_N} \int_{y_1 \in \mathcal{R}^{i_1}} \ldots
  \int_{y_{N} \in \mathcal{R}^{i_N}} \left\{\mathbb{E}[\|\mathbf{X} -
    \widehat{\mathbf{X}}(\mathbf{I}_1^N) \|_2^2 |
    \mathbf{Y=y},\mathbf{I}_1^N=\mathbf{i}_1^N] \right\}f(\mathbf{y})
  d \mathbf{y}&
\end{aligned}
\end{equation}
\setcounter{equation}{\value{MYtempeqncnt}}
\hrulefill
\end{figure*}
\setcounter{equation}{6}

Then, using the fixed codebooks $\mathcal{C}_n$ $(n=1,\ldots,N)$, the
MSE-minimizing encoding indexes are identical to finding the ones
which minimize the term in the braces in the last expression of
\eqref{eq:suboptimal condition} since $f(\mathbf{y})$ is always non-negative.
The resulting optimal indexes denoted by $\{i_n^\star \in \mathcal{I}\}_{n=1}^N$ are obtained by \eqref{eq:proof enc}, where $(a)$ holds since the conditional expectation of sum of RV's equals to the sum of their conditional expectations. $(b)$ follows from the fact that $\mathbf{X}$ is independent of $\mathbf{I}_1^N$, conditioned on $\mathbf{Y}$, and hence, $\mathbb{E} \left[\|\mathbf{X}\|_2^2 |  \mathbf{Y\!=\!y}, \mathbf{I}_1^N\!=\!\mathbf{i}_1^N \right] = \mathbb{E} \left[\|\mathbf{X}\|_2^2 | \mathbf{Y\!=\!y}\right]$ which
is pulled out of the optimization. Also, $(c)$ follows from a similar
rationale, i.e., $\widehat{\mathbf{X}}(\mathbf{I}_1^N)$ is independent
of $\mathbf{Y}$, conditioned on $\mathbf{I}_1^N$. Further, $\mathbf{X}$
and $\widehat{\mathbf{X}}(\mathbf{I}_1^N)$ are independent conditioned
on $\mathbf{Y}$ and $\mathbf{I}_1^N$.
\begin{figure*}[!t]
\normalsize
\setcounter{MYtempeqncnt}{\value{equation}}
\begin{equation} \label{eq:proof enc}
\begin{aligned}
    \{i_n^\star\}_{n=1}^N &= \textrm{arg }\underset{\mathbf{i}_1^N}{\textrm{min }} \mathbb{E}[\|\mathbf{X} - \widehat{\mathbf{X}}(\mathbf{I}_1^N) \|_2^2 | \mathbf{Y=y},\mathbf{I}_1^N=\mathbf{i}_1^N]& \\
    &\stackrel{(a)}{=} \textrm{arg }\underset{\mathbf{i}_1^N}{\textrm{min }} \left\{\mathbb{E} \left[\|\mathbf{X}\|_2^2 | \mathbf{Y\!=\!y}, \mathbf{I}_1^N=\mathbf{i}_1^N  \right] + \mathbb{E}\left[\|\widehat{\mathbf{X}}(\mathbf{I}_1^N) \|_2^2 | \mathbf{Y=y} , \mathbf{I}_1^N=\mathbf{i}_1^N\right] - 2\mathbb{E}\left[\mathbf{X}^T \widehat{\mathbf{X}}(\mathbf{I}_1^N)  | \mathbf{Y=y}, \mathbf{I}_1^N=\mathbf{i}_1^N\right] \right\}& \\
    &\stackrel{(b)}{=} \textrm{arg }\underset{\mathbf{i}_1^N }{\textrm{min }} \left\{ \mathbb{E}[\|\widehat{\mathbf{X}}(\mathbf{I}_1^N)\|_2^2  | \mathbf{Y\!=\!y}, \mathbf{I}_1^N=\mathbf{i}_1^N] - 2\mathbb{E}[\mathbf{X}^T \widehat{\mathbf{X}}(\mathbf{I}_1^N)  | \mathbf{Y=y}, \mathbf{I}_1^N=\mathbf{i}_1^N] \right\}& \\
    &\stackrel{(c)}{=} \textrm{arg }\underset{\mathbf{i}_1^N}{\textrm{min }} \left\{\mathbb{E}[\|\widehat{\mathbf{X}}(\mathbf{I}_1^N)\|_2^2 \big | \mathbf{I}_1^N=\mathbf{i}_1^N] - 2\mathbb{E}[\mathbf{X}^T \big | \mathbf{Y=y}] \mathbb{E}[\widehat{\mathbf{X}}(\mathbf{I}_1^N) \big | \mathbf{I}_1^N=\mathbf{i}_1^N] \right\}&
\end{aligned}
\end{equation}
\setcounter{equation}{\value{MYtempeqncnt}}
\hrulefill
\end{figure*}
\setcounter{equation}{7}
For a realization $\mathbf{Y=y}$, the last expression in \eqref{eq:proof enc} can be rewritten as (using \eqref{eq:MMSE est})
\begin{equation} \label{eq:final enc original}
\begin{aligned}
    \{i_1^\star,\ldots,i_N^\star\} &= \textrm{arg }\underset{\mathbf{i}_1^N}{\textrm{min}} \left\{\|\widehat{\mathbf{x}}(\mathbf{i}_1^N) \|_2^2- 2 \widetilde{\mathbf{x}}(\mathbf{y})^{T} \widehat{\mathbf{x}}(\mathbf{i}_1^N)\right\},&
\end{aligned}
\end{equation}
where $\mathbf{i}_1^N$ denotes the sequence $i_1,\ldots,i_N$.

Unfortunately, solving \eqref{eq:final enc original} jointly for all
encoding indexes is not analytically and practically tractable for a
general sparse reconstruction function as it requires reconstruction and searching over
all possible codepoints, leading to prohibitive complexity. Since we are
interested in developing a reasonably simple coding system, we refrain from
the optimal joint encoder. Instead, in this work, we focus on
sub-optimal techniques (with respect to \eqref{eq:final enc original}) for
quantization of CS measurements. The quantization schemes are
developed in the next section.

\section{Analysis-by-Synthesis Quantization of CS Measurements} \label{sec:quant}

\begin{figure}
\begin{center}
  \psfrag{x}[][][0.65]{$\mathbf{X}$}
  \psfrag{Psi}[][][0.8]{$\mathbf{\Phi}$}
  \psfrag{y}[][][0.6]{$\mathbf{Y}$}
  \psfrag{E}[][][0.7]{Quantizer encoder}
  \psfrag{CS_E}[][][0.7]{CS encoder}
  \psfrag{D}[][][0.7]{Quantizer decoder}
  \psfrag{Q}[][][0.75]{$\textsf{E}$}
  \psfrag{C_1}[][][0.65]{$\mathcal{C}_1$}
  \psfrag{C_N}[][][0.65]{$\mathcal{C}_N$}
  \psfrag{I_1}[][][0.6]{$I_1$}
  \psfrag{I_N}[][][0.6]{$I_N$}
  \psfrag{y_1}[][][0.6]{$Y_1$}
  \psfrag{y_N}[][][0.6]{$Y_N$}
  \psfrag{Q-1}[][][0.75]{$\textsf{D}$}
  \psfrag{y_h_1}[][][0.6]{$\widehat{Y}_1$}
  \psfrag{y_h_N}[][][0.6]{$\widehat{Y}_N$}
  \psfrag{y_h}[][][0.6]{$\widehat{\mathbf{Y}}$}
  \psfrag{Rec}[][][0.65]{CS decoder}
  \psfrag{R}[][][0.8]{\textsf{R}}
  \psfrag{x_h}[][][0.65]{$\widehat{\mathbf{X}}$}
  \includegraphics[width=9cm]{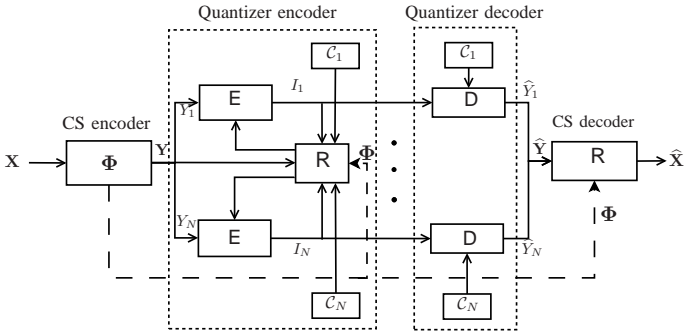}\\
  \caption{Analysis-by-synthesis (AbS) quantization of CS measurements. At the quantizer encoder, the function \textsf{R} performs as a \textit{synthesizer} based on a sparse reconstruction, and the function \textsf{E} acts as an \textit{analyzer} based on the optimization problem \eqref{eq:final enc} in a closed-loop.}
  \label{fig:diagram_CS}
  \vspace{-0.5cm}
  \end{center}
\end{figure}

In this section, we first show how an encoding index can be chosen by
fixing the other indexes under the assumptions of \probref{prob1}, and then
develop AbS-based quantization schemes following by complexity
analysis. Our studied AbS system is illustrated in
\figref{fig:diagram_CS}.

In order to solve \eqref{eq:final enc original} approximately, we
consider optimizing one variable by fixing the others, that is,
optimizing the index $i_n$ by fixing the indexes
$i_1,\ldots,i_{n-1},i_{n+1},\ldots,i_{N}$. This is indeed an
\textit{alternating} optimization approach and sub-optimal
compared to the joint optimization method \eqref{eq:final enc
  original}, but provides a feasible solution with low complexity.

Now, assume that given the codebook sets, we fix all encoding indexes except $i_n$ ($n \in \{1,\ldots,N\}$). Then, from \eqref{eq:final enc original}, the encoder attains the following encoding rule
\begin{equation} \label{eq:final enc}
\begin{aligned}
    i_n^\star &= \textrm{arg }\underset{i_n \in \mathcal{I}}{\textrm{min}} \left\{\|\widehat{\mathbf{x}}(i_n) \|_2^2- 2 \widetilde{\mathbf{x}}(\mathbf{y})^{T} \widehat{\mathbf{x}}(i_n)\right\}.&
\end{aligned}
\end{equation}
By denoting $\widehat{\mathbf{x}}(i_n) \! \triangleq \!
\textsf{R}\left([c_{i_1},\ldots,c_{i_n},\ldots,c_{i_N}\right)]^T$, we
refer that the reconstructed signal is dependent only upon the index
(codepoints) associated with the $n^{th}$ measurement
entry. Interestingly, \eqref{eq:final enc} indicates that in order to
minimize the end-to-end MSE for a fixed codebook, the index $i_n$
should be chosen such that the final reconstruction vector is as close
as possible (in $\ell_2$-norm) to the MMSE estimation of the sparse
source given the measurements. We employ \eqref{eq:final enc} in an
iterate-alternate optimization approach to realize new AbS-based
algorithms which are described in the next subsection. Before going to the next subsection, we have the following remark.

\begin{rem} \label{rem:MMSE estimator} As shown in our formulations
  (e.g. \eqref{eq:final enc}), we need the MMSE estimate
  $\widetilde{\mathbf{x}}(\mathbf{y})$ to find the encoding
  indexes. However, in practice, implementing the MMSE estimator is
  not feasible. Therefore, we approximate the MMSE estimate $\widetilde{\mathbf{x}}(\mathbf{y})$ by the sparse signal estimate of reconstruction function \textsf{R} using unquantized measurement $\mathbf{y}$. Denoting the sparse signal estimate by $\bar{\mathbf{x}}(\mathbf{y})$, we assume $\widetilde{\mathbf{x}}(\mathbf{y}) \approx \bar{\mathbf{x}}(\mathbf{y})$. In this paper, we use the
  greedy \textit{orthogonal matching pursuit} (OMP)
  \cite{07:Tropp,08:Blumensath} reconstruction algorithm as the sparse reconstruction function \textsf{R}. The choice of OMP is motivated due to a good tradeoff between complexity and reconstruction performance. The OMP algorithm is briefly described in the Appendix. We emphasize that our formulation is general and does not deter use of any practical reconstruction algorithm in lieu of OMP.
\end{rem}

\subsection{Proposed Quantization Algorithms} \label{subsec:algorithm}
We first describe the \textit{iterative} framework for the proposed
quantization schemes summarized in \algref{alg1}. Suppose that the
codebook sets $\mathcal{C}_n$ ($n=1,\ldots,N$) are designed offline,
and let the quantizer encoder have access to the sensing matrix
$\mathbf{\Phi}$ as well as the codebooks (step
(1)). \footnote[1]{Note that the sparsity level $K$ may be also provided at the encoder and the decoder if the OMP algorithm is used as the reconstruction function \textsf{R}. If subspace pursuit \cite{09:Dai} or CoSaMP \cite{09:Needell} reconstruction algorithms are used, then $K$ must be provided. However, using the basis pursuit \cite{98:Chen} or LASSO \cite{96:Tibshirani} reconstruction algorithms, the sparsity level is not necessarily required.} In step (2), we obtain the locally reconstructed vector $\bar{\mathbf{x}}(\mathbf{y})$ as an approximation to the MMSE estimator $\widetilde{\mathbf{x}}(\mathbf{y})$. Now, we define a dummy vector $\mathbf{z} \in \mathbb{R}^N$, where its $n^{th}$ component is chosen uniformly at random from the set $\mathcal{C}_{n}$ ($\forall n$) at the first iteration (step (3)). The vector $\mathbf{z}$ stores coefficients of quantized CS measurements. Throughout iterations, the entries of $\mathbf{z}$ are either \textit{sequentially} or \textit{non-sequentially} adjusted for minimizing the reconstruction MSE. Now, we describe the subroutine $\texttt{AbS}(\cdot)$ executed in \algref{alg1} (step (6)) which uses two different \textit{alternating} approaches (in terms of
performance and complexity) for implementing \eqref{eq:final enc}.

\begin{algorithm}
\caption{: AbS quantization}\label{alg1}
\begin{algorithmic}[1]
\STATE{\textbf{input: } $\mathcal{C}_n = \{c_{i_n}\}_{i_n=0}^{2^{r_y}-1}$ ($\forall n=1,\ldots,N$) and $\mathbf{\Phi , y}$, $\gamma$ (stopping threshold)}
\STATE{\textbf{compute: } $\bar{\mathbf{x}}(\mathbf{y})$ (locally reconstructed vector)}
\STATE{\textbf{initialize } $\mathbf{z}^{(0)} \in \mathbb{R}^N$, where $z_n^{(0)} \in \mathcal{C}_{n},$ $\forall n$.}
\STATE{Set $l \gets 0$ (iteration counter)}
\REPEAT
    \STATE{$[i_n^\star , \widehat{\mathbf{x}}^{(l+1)}{(i_n^\star)},\mathbf{z}^{(l+1)}] = \texttt{AbS}(\mathcal{C}_n,\bar{\mathbf{x}}(\mathbf{y}),\mathbf{z}^{(l)})$ , $\forall n$}
    \STATE{$l \gets l+1$}
\UNTIL{$\left| \left[ \|\widehat{\mathbf{x}}^{(l)}(i_n^\star) \|_2^2- 2 \bar{\mathbf{x}}(\mathbf{y})^{T} \widehat{\mathbf{x}}^{(l)}(i_n^\star)  \right] \right.$ \\
$\hspace{1cm}\left.- \left[ \|\widehat{\mathbf{x}}^{(l-1)}(i_n^\star) \|_2^2 + 2 \bar{\mathbf{x}}(\mathbf{y})^{T} \widehat{\mathbf{x}}^{(l-1)}(i_n^\star) \right] \right| <  \gamma$ , $\forall n$}
\STATE{\textbf{output: } $I_n = i_n^{\star}$ , $\widehat{Y}_n = c_{i_n^{\star}}$ , $\forall n$ }
\end{algorithmic}
\end{algorithm}

\begin{enumerate}

\item \textbf{Sequential AbS quantization:} Our first AbS-based quantization scheme which can be executed in \algref{alg1} is summarized in the Subroutine $\texttt{AbS\_seq} (\cdot)$ where the main idea is that each measurement entry is sequentially adjusted towards the direction of its MSE-minimizing codepoint at each iteration.

\vspace{0.1cm}
\rule{\linewidth}{1pt}
$\textbf{Subroutine: } \texttt{AbS\_seq} \left(\mathcal{C}_n , \bar{\mathbf{x}}(\mathbf{y}),\mathbf{z}^{(l)}\right)$
\rule{\linewidth}{0.4pt}
\begin{algorithmic}[1]
    \FOR{$n=1:N$}
        \FOR{$i=0:2^{r_y}-1$}
           \STATE{$z_n^{(l)} \gets c_{i_n}$}
            \STATE{\textbf{compute:} $\widehat{\mathbf{x}}^{(l)}(i_n) = \textsf{R} (\mathbf{z}^{(l)})$}
        \ENDFOR
        \STATE{$i_n^{\star} =  \underset{i_n \in \mathcal{I}}{\textrm{arg min}} \{\|\widehat{\mathbf{x}}^{(l)}(i_n)\|_2^2-2\bar{\mathbf{x}}(\mathbf{y})^{T} \widehat{\mathbf{x}}^{(l)}(i_n)\}$}
        \STATE{\textbf{update:} $z_n^{(l)} \gets c_{i_n^{\star}}(i_n)$}
    \ENDFOR
\STATE{$\textbf{output: }  i_n^\star , \widehat{\mathbf{x}}^{(l)}(i_n^\star) ,  \mathbf{z}^{(l)}$}
\hspace*{-0.6cm}\rule{1.08\linewidth}{0.4pt}
\end{algorithmic}

Using \algref{alg1}, the function $\texttt{AbS\_seq}(\cdot)$ accepts the codebooks $\mathcal{C}_n$, $\forall n$, the locally reconstructed vector $\bar{\mathbf{x}}(\mathbf{y})$ and the dummy vector $\mathbf{z}^{(l)}$. At iteration $l$, the $n^{th}$ ($n=1,\ldots,N$) entry of $\mathbf{z}^{(l)}$, denoted by $z_n^{(l)}$, is replaced by all $2^{r_y}$ codepoints from the set $\mathcal{C}_n$ (step (3)) while the other entries are fixed, and the reconstructed vectors, denoted by $\widehat{\mathbf{x}}^{(l)}(i_n) = \textsf{R} (\mathbf{z}^{(l)})$ ($i_n \in \mathcal{I}=\{0,\ldots,2^{r_y}-1\}$), are synthesized corresponding to each vector (step (4)). Then, an optimization is carried out by solving $\underset{i_n \in \mathcal{I}}{\textrm{arg min}} \{\|\widehat{\mathbf{x}}^{(l)}(i_n)\|_2^2-2\bar{\mathbf{x}}(\mathbf{y})^{T} \widehat{\mathbf{x}}^{(l)}(i_n)\}$ so as to find the wining MSE-minimizing encoding index $i_n^\star$ (step (6)). Next, the $n^{th}$ entry of the vector $\mathbf{z}^{(l)}$ is updated by the codepoint associated with the analyzed index, i.e., $c_{i_n^{\star}}$ (step (7)). This procedure continues for each entry of $\mathbf{z}^{(l)}$ sequentially, and the subroutine produces the optimized encoding index $i_n^\star$, and the reconstructed vector $\widehat{\mathbf{x}}^{(l)}(i_n^\star)$ as well as the updated quantized vector $\mathbf{z}^{(l)}$ (step (9)) which will be used by the subroutine at the next iteration of \algref{alg1}.

\item \textbf{Non-sequential AbS quantization:} Using \algref{alg1}, our second proposed Subroutine $\texttt{AbS\_nonseq}(\cdot)$ accepts the codebooks $\mathcal{C}_n$, $\forall n$, the locally reconstructed vector $\bar{\mathbf{x}}(\mathbf{y})$ and the dummy vector $\mathbf{z}^{(l)}$. This non-sequential scheme is not order-dependent, and finds the MSE-minimizing index/codepoint by tracking the best path at each iteration.

\rule{\linewidth}{1pt}
$\textbf{Subroutine: } \texttt{AbS\_nonseq} \left(\mathcal{C}_n , \bar{\mathbf{x}}(\mathbf{y}),\mathbf{z}^{(l)}\right)$
\rule{\linewidth}{0.4pt}
\begin{algorithmic}[1]
    \STATE{$\mathcal{L} = \varnothing$}
\REPEAT
    \STATE{\textbf{initialize:} $\mathbf{\Delta} = \mathbf{0}^{({N-  |\mathcal{L}|}) \times 2^{r_y}}$}
    \FOR{$n \in \{1,\ldots,N\}  {\tt\char`\\}  \mathcal{L}$}
        \FOR{$i=0:2^{r_y}-1$}
           \STATE{$z_n^{(l)} \gets c_{i_n}$}
            \STATE{\textbf{compute:} $\widehat{\mathbf{x}}^{(l)}(i_n) = \textsf{R} (\mathbf{z}^{(l)})$}
        \ENDFOR
        \STATE{\textbf{compute:} $\mathbf{\Delta}(n,i\!\!+\!\!1) \!\!=\!\! \|\widehat{\mathbf{x}}^{(l)}(i_n)\|_2^2 \!-\! 2 \bar{\mathbf{x}}(\mathbf{y})^{T} \widehat{\mathbf{x}}^{(l)}(i_n)$}
    \ENDFOR
    \STATE{$[n^\star , i_n^{\star}] =  \underset{n,i}{\textrm{arg min }} \mathbf{\Delta}(n,i\!+\!1)$}
    \STATE{\textbf{update:} $z_{n^\star}^{(l)} \gets c_{i_n^{\star}}$}
    \STATE{$\mathcal{L} \gets \mathcal{L} \cup \{n^\star\}$}
\UNTIL{$\mathcal{L}=\{1,2,\ldots,N\}$}
\STATE{$\textbf{output: }  i_n^\star , \widehat{\mathbf{x}}^{(l)}(i_n^\star) ,  \mathbf{z}^{(l)}$}
\hspace*{-0.6cm}\rule{1.08\linewidth}{0.4pt}
\end{algorithmic}

    At the first iteration of $\texttt{AbS\_nonseq}(\cdot)$, we define a set $\mathcal{L}$ which is initially empty (step (1)). At iteration $l$ of the algorithm, the $n^{th}$ ($n=1,\ldots,N$) entry of $\mathbf{z}^{(l)}$, denoted by $z_n^{(l)}$, is replaced by all $2^{r_y}$ codepoints in the set $\mathcal{C}_n$ (step (6)) while the other entries are fixed, and the reconstructed vectors, denoted by $\widehat{\mathbf{x}}^{(l)}(i_n) = \textsf{R} (\mathbf{z}^{(l)})$ ($i_n \in \mathcal{I}=\{0,\ldots,2^{r_y}-1\}$), are synthesized corresponding to each vector (step (7)) whose values are stored in the matrix $\mathbf{\Delta} \in \mathbb{R}^{({N-  |\mathcal{L}|}) \times 2^{r_y}}$ (step (9)). Then, an analysis is performed by solving \eqref{eq:final enc}, i.e., $\underset{n,i}{\textrm{arg min}} \{\|\widehat{\mathbf{x}}^{(l)}(i_n)\|_2^2- 2 \bar{\mathbf{x}}(\mathbf{y})^{T} \widehat{\mathbf{x}}^{(l)}(i_n)\}$, through a search in rows and columns of the matrix $\mathbf{\Delta}$, so as to find the MSE-minimizing encoding index $i_n^\star$ and the entry's index $n^\star$ (step (11)). Now, the entry $n^{\star}$ of the vector $\mathbf{z}^{(l)}$ is updated by the codepoint associated with the analyzed index, i.e., $c_{i_n^{\star}}$ (step (12)). Note that the set $\mathcal{L}$ expands by adding $n^\star$ to the previous set (step (13)) in order to exclude the minimizing index and entry at the next iteration of the subroutine. The iterations continue until the set $\mathcal{L}$ consists of all elements of the set $\{1,2,\ldots,N\}$ so that all entries are assigned by the minimizing codepoints (step (14)). At the last iteration of the subroutine, it outputs the optimized transmission index $i_n^\star$, and the reconstructed vector $\widehat{\mathbf{x}}^{(l)}(i_n^\star)$ as well as the updated quantized vector $\mathbf{z}^{(l)}$ (step (15)) which will be used by the subroutine at the next iteration of \algref{alg1}.

\end{enumerate}

\vspace{0.2cm}

\algref{alg1} iterates until convergence where the stopping criterion is that reconstruction improvement at two consecutive iterations is smaller than a predefined threshold $\gamma > 0$. After convergence, the algorithm outputs the quantization indexes $I_n$'s and the CS quantized measurements $\widehat{Y}_n$'s, $n=1,\ldots,N$, (step (9)) in which the latter is regarded as an input to the final sparse reconstruction algorithm in order to provide the estimate $\widehat{\mathbf{x}}$. In the following, we discuss the convergence of \algref{alg1}.

\begin{rem} \label{rem:convergence}
Following standard convergence proofs in \cite[Lemmas 11.3.1-2]{91:Gersho}, we provide qualitative arguments on the convergence of Algorithm 1. By construction (and ignoring issues such as numerical precision), the iterative design in Algorithm 1 given codebook sets $\mathcal{C}_n$ $(n=1,\ldots,N)$ and a fixed sparse reconstruction algorithm $\textsf{R}$ always converges to a local optimum. More precisely, at each iteration of \algref{alg1}, given the fixed codebook sets and a sparse reconstruction algorithm, whenever the criteria in step (6) of $\texttt{AbS\_seq}$ or step (11) of $\texttt{AbS\_nonseq}$ are invoked, the reconstruction distortion given the updated index and the remaining fixed indexes can only leave unchanged or reduced. This is due to the fact that the distortion-minimizing index is always chosen. Hence, the distortion monotonically decreases at each iteration. Since the stopping criterion is defined as the difference in distortion at successive iterations, the stopping condition is bound to satisfy after finite number of iterations, and the algorithm converges to optimized encoding indexes. However, nothing can be generally guaranteed about global optimality.
\end{rem}

\subsection{Complexity Analysis} \label{subsec:complexity}
In this section, we analyze the encoding computational complexity of the proposed quantization schemes. We mainly quantify how many times a sparse reconstruction algorithm is invoked throughout the AbS procedures. Note that, in practice, other operators used in the proposed algorithms have negligible complexity compared to the sparse reconstruction algorithms.

First, recall from \eqref{eq:final enc original} that an exhaustive search for the joint optimization requires $\mathcal{O}(2^{Mr_x})$, or $\mathcal{O}(2^{Nr_y})$ (since $M r_x = N r_y$), computations of a sparse reconstruction algorithm. This is not feasible in practice. Employing the sequential AbS quantization (Subroutine $\texttt{AbS\_seq}$), the operations for calculating the encoding indexes increase at most like $\mathcal{O}(N 2^{\frac{Mr_x}{N}})$ at each iteration of \algref{alg1}, where it follows that at $L$ iterations, the total complexity increases with $\mathcal{O}(L N 2^{\frac{Mr_x}{N}})$. Next, let us consider the non-sequential AbS-based quantization (Subroutine $\texttt{AbS\_nonseq}$) at one iteration of \algref{alg1}. The operations for calculating encoding indexes require $N2^{Mr_x/N} + (N-1)2^{Mr_x/N} + \ldots + 2^{Mr_x/N} = \frac{N(N+1)}{2} 2^{Mr_x/N}$ computations of a sparse reconstruction algorithm since at each iteration of the subroutine the set $\mathcal{L}$ expands by adding one element. Hence, it can be shown that the computational complexity of the non-sequential AbS quantization after $L$ iterations of \algref{alg1} grows at most like $\mathcal{O}(L \frac{N^2}{2} 2^{\frac{Mr_x}{N}})$.

\begin{figure}
  \begin{center}
  \includegraphics[width=\linewidth,height=7.5cm]{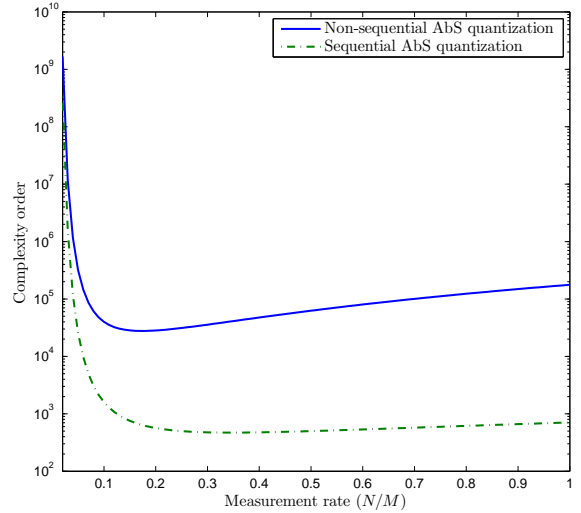}\\
  \caption{Complexity order as a function of measurement rate $\frac{N}{M}$ for the proposed AbS quantization schemes for $M=500$ and $r_x=0.5$ bit/component. }
  \label{fig:complexity}
  \end{center}
\end{figure}

The order of computations for the AbS schemes indicates a tradeoff between availability of compression resources, i.e., number of measurements and quantization bit budget. It also implies that, for a fixed bit budget $R_x = Mr_x$, by increasing the number of measurements, first the complexity decays sharply, and then at some point it starts increasing with a small slope. This is due to the fact that the complexity depends on the number of measurements (through the linear term $N$ or the quadratic term $N^2$ for the sequential and the non-sequential AbS-based algorithms, respectively) and the quantization bit-rate per measurement entry $r_y=Mr_x/N$ (through the exponential term $2^{\frac{Mr_x}{N}}$). This aspect is shown in \figref{fig:complexity}.

We finalize this section with a remark regarding codebook design for the studied quantization schemes.

\begin{rem} \label{rem:codebook} As mentioned in the design of the
  proposed schemes, we have assumed that the codebook sets are
  given. However, using unoptimized codebooks may lead to poor
  performance. Here, we describe an alternative to training codebooks
  corresponding to each quantization scheme. A standard approach would
  be the \textit{Lloyd algorithm} \cite[Chapter 6]{91:Gersho}, where,
  with possibly random initializations, either the encoder (index
  allocation) or decoder (codepoints) is assumed known and the other
  is selected optimally with respect to minimizing a particular
  distortion measure and a statistically specified input. This
  procedure is then alternated and iterated until (local) convergence
  is reached.

  Due to the discrete mixture distribution of sparse source and highly
  non-linear behavior of sparse reconstruction algorithms, it is generally
  very challenging to derive closed-form optimal codepoints with
  respect to minimizing the end-to-end MSE. Therefore, we devise a
  potentially sub-optimal design for codebook training. Let us assume
  that CS measurements are identically and independently distributed
  (i.i.d.), and introduce the average quantization distortion
  $\mathbb{E}[|Y_n - \widehat{Y}_n|^2]$ as a criterion. Then, all the
  codebooks are the same, and the codepoints are optimized using the
  Lloyd algorithm for a random measurement entry $Y_n$. \footnote[1]{Note that in the case of exactly $K$-sparse vector whose non-zero components are i.i.d. RV's, using the central limit theorem, each entry of the linear measurement vector converges weakly to a Gaussian random variable with zero mean and variance $K/N$ as $(K,M,N) \rightarrow \infty$ and with the rates $K/M$ and $N/M$ remain constant (see \cite{11:Dai} for more details).} The fact that all codebooks are chosen the same for each measurement entry can be justified by similar observations in \cite{11:Laska}. The optimized codepoints using the Lloyd algorithm would minimize the quantization distortion per measurement entry. Therefore, these codepoints can be considered \textit{good} alternatives for the AbS-based quantizers which take the end-to-end distortion into account. The codebook sets are designed offline, and will be used for implementation of the quantization schemes whose performances are evaluated in \secref{sec:numerical}.

\end{rem}

In what follows, we investigate coding of the signal domain where its combination with coding of the CS measurement domain leads to an adaptive design algorithm.

\section{Adaptive Coding: Signal and Measurement Domains} \label{subsec:adaptive coding}
Up to this point, we have developed new schemes for quantization in
the CS measurement domain. Since using the proposed AbS designs, we
reconstruct the source vector according to which the
quantization is performed, it is important to investigate a scenario
where the locally reconstructed source $\bar{\mathbf{x}}$ (we drop the dependency of the vector on $\mathbf{y}$ for simplicity of notation) is coded.
\subsection{Signal Domain Coding}
In order to efficiently address the design problem which aims to quantize in the domain of reconstructed signal, two alternatives may be visualized. They are as follows.

\begin{enumerate}
    \item \textbf{Direct coding:} Using this scheme, all coefficients of the vector $\bar{\mathbf{x}} \in \mathbb{R}^M$ are quantized with available $r_x$ bits per component. Therefore, in order to encode all coefficients, this scheme requires $R_x \geq M$ (or, $r_x \geq 1$) bits. Formally, let the codebook sets $\mathcal{G}_m \triangleq \{g_{i_m} \in \mathbb{R}\}_{i_m=0}^{2^{r_x}-1}$ corresponding to components $\bar{x}_m$'s ($m=1,\ldots,M$) be given. Then, the encoding index associated with the component $\bar{x}_m$ is chosen as
        \begin{equation} \label{eq:direct}
            i_m^\star = \underset{i_m \in \mathcal{J}}{\textrm{arg min }} |\bar{x}_m - g_{i_m}|^2, \hspace{0.25cm} m=1,\ldots, M ,
        \end{equation}
        where the index set $\mathcal{J}$ is defined as $\mathcal{J} \triangleq \{0,\ldots,2^{r_x}-1\}$. The decoder functions according to a look-up table $I_m = i_m \Rightarrow \widehat{X}_m = g_{i_m}$. Using the direct coding scheme, all the codebooks are assumed the same (since the components of $\mathbf{X}$ are i.i.d. RV's), and codepoints are optimized using the Lloyd algorithm by applying the performance measure $\mathbb{E}[|X_m - \widehat{X}_m|_2^2]$.

      \item \textbf{Support set coding:} In this case, we take into
        account the sparsity pattern of the locally reconstructed vector
        $\bar{\mathbf{x}}$. We first code the reconstructed support
        set, and then the magnitude of the non-zero coefficients on
        the support set. We denote by $\bar{\mathbf{x}}_{\widehat{\mathcal{S}}} \in \mathbb{R}^K$
        the entries of $\bar{\mathbf{x}}$ indexed by the elements of
        the estimated support set, denoted by $\widehat{\mathcal{S}}
        \subset \{1,\ldots,M\}$. Therefore, each component of
        $\widehat{\mathcal{S}}$ can be represented by $\log_2 M$ bits
        that can be coded, and then recovered without loss. Now, we
        map the magnitude of $K$ largest non-zero coefficients of
        $\bar{\mathbf{x}}$ to their nearest codepoints using $R_0
        \triangleq R_x - K \log_2 M$ bits, where $R_x = M r_x$ is the
        total bit budget. Suppose the codebooks corresponding to the
        estimated non-zero coefficients, denoted by $\bar{x}_s$, $s
        \in \widehat{\mathcal{S}}$, be given by $\mathcal{D} =
        \{d_{i_s} \in \mathbb{R}\}_{i_s=0}^{2^{R_0/K}-1}$. Then, the
        encoding index associated with entry $\bar{x}_{s}$ is chosen
        as
        \begin{equation} \label{eq:ss}
            i_s^\star = \underset{i_s \in \mathcal{K}}{\textrm{arg min }} |\bar{x}_s - d_{i_s}|^2, \hspace{0.25cm} s \in \widehat{\mathcal{S}},
        \end{equation}
        where the index set $\mathcal{K}$ is defined as $\mathcal{K}
        \triangleq \{0,\ldots,2^{R_0/K}-1\}$. The decoder works
        according to a look-up table $I_s = i_s \Rightarrow
        \widehat{X}_s = d_{i_s}$. For quantizing all non-zero
        coefficients, this approach requires at least $R_x \geq K \log_2 M + K$ bits.

        Looking at the direct and support set coding, it can be
        inferred that these approaches require rather high bit budget
        for quantization. Exploiting the support set coding approach, the Lloyd algorithm is used for codebook training by adopting the performance criterion $\mathbb{E}[|X_s - \widehat{X}_s|_2^2]$ for the input RV $X_s$, $s \in \mathcal{S}$, where $X_s$ is a non-zero coefficient of $\mathbf{X}$ drawn according to a known distribution.

\end{enumerate}

Next, we show how to combine the signal domain coding schemes with the CS measurement domain quantization schemes in order to adaptively gain a better performance.

\subsection{Adaptive Coding}
        Till now, we have proposed quantization schemes in CS measurement domain and signal domain.
        An engineering approach is to choose adaptively a better quantized signal (when it is
        compared with the locally reconstructed signal
        $\bar{\mathbf{x}}$) between the signal domain coding and the
        CS measurement domain coding schemes. This can be performed by
        assigning 2 flag bits for representing the proposed four schemes at
        the decoder, i.e.,
        \begin{enumerate}
            \item nearest-neighbor coding (CS measurement domain),
            \item non-sequential AbS quantization (CS measurement domain),
            \item direct coding (signal domain), and
            \item support set coding (signal domain).
        \end{enumerate}
        Then, the remaining $Mr_x-2$ bits are used for quantization of each individual scheme. 
        Since the computational complexity of the direct, support set and nearest-neighbor coding schemes are negligible compared to the non-sequential AbS quantization scheme, the total complexity of the adaptive coding grows at most as that of the non-sequential AbS quantization.

\section{Experiments and Results} \label{sec:numerical}
In this section, we first demonstrate the performance of the proposed quantization schemes:
\begin{enumerate}
    \item Nearest-neighbor coding (for quantizing in the CS measurement domain),
    \item sequential AbS quantization (for quantizing in the CS measurement domain),
    \item non-sequential AbS quantization (for quantizing in the CS measurement domain),
    \item direct coding (for quantizing in the signal domain),
    \item support set coding (for quantizing in the signal domain), and
    \item adaptive coding (for quantizing in both domains).
\end{enumerate}
Finally, we compare the performance of AbS-based quantizer vis-a-vis existing methods in the literature, specifically, the methods from \cite{09:Sun}. \footnote[1]{In the spirit of reproducible results, we provide MATLAB codes for simulation of the AbS-based quantizers in the following website: www.ee.kth.se/$\sim$amishi/reproducible$\_$research.html.}

We mainly quantify the MSE obtained by these schemes in terms of availability of compression resources, i.e., number of measurements and quantization bit-rate. Before showing simulation results, we state experimental setups in the next subsection.

\subsection{Experimental Setups} \label{subsec:setup}
We quantify the performance using normalized MSE (NMSE) defined as
\begin{equation} \label{eq:NMSE}
    \textrm{NMSE} \triangleq \frac{\mathbb{E}[\|\mathbf{X}-\widehat{\mathbf{X}}\|_2^2]}{\mathbb{E}[\|\mathbf{X}\|_2^2]}.
\end{equation}
In principle, the numerator of NMSE in \eqref{eq:NMSE} is computed by sample averaging over generated realizations of $\mathbf{X}$ using Monte-Carlo simulations, and the denominator can be calculated exactly under the assumptions in our simulation setup. This calculation will be given in details later.

In addition, in order to measure the level of under-sampling, we define measurement rate ($0 < \alpha \leq 1$) as
\begin{equation} \label{eq:fom}
    \alpha \triangleq N/M.
\end{equation}
Our simulation setup includes the following steps:
\begin{enumerate}
  \item For given values of sparsity level $K$ (assumed known in advance) and input vector size $M$, choose $\alpha$, and round the number of measurements $N$ to its nearest integer.
  \item Randomly generate a set of exactly $K$-sparse vector $\mathbf{X}$ where the support set $\mathcal{S}$ with $|\mathcal{S}| = K$ is chosen uniformly at random over the set $\{1,2,\ldots,M\}$. The non-zero coefficients of $\mathbf{X}$ are i.i.d. R.V.'s drawn from a known distribution. In our simulations, we use two mostly-common distributions for the non-zero coefficients: Gaussian and uniform. Based on the uniform sparsity pattern assumption, $\mathbb{E}[\|\mathbf{X}\|_2^2]$ (the denominator in \eqref{eq:NMSE}) can be analytically derived. It follows that
    \begin{equation}
    \begin{aligned}
        &\mathbb{E}[\|\mathbf{X}\|_2^2] = \sum_{m=1}^M \mathbb{E}[X_m^2]& \\
        &= \sum_{m=1}^M \textrm{Pr}(m \!\in \!\mathcal{S}) \mathbb{E}[X_m^2 | m \! \in \! \mathcal{S}] + \textrm{Pr}(m \! \notin \!\mathcal{S}) \mathbb{E}[X_m^2 | m \! \notin \! \mathcal{S}]& \\
        &\stackrel{(a)}{=} \sum_{m=1}^M \frac{K}{M} \mathbb{E}[X_m^2 | m \! \in \! \mathcal{S}]&
    \end{aligned}
    \end{equation}
    where $(a)$ follows from the assumption of the uniformly distributed sparsity pattern, and also from the fact the second moments of the coefficients of $\mathbf{X}$ that are not within the support set $\mathcal{S}$ are zero. Now, note that $\mathbb{E}[X_m^2 | m \! \in \! \mathcal{S}]$ shows the second moment of $X_m$ within the support set $\mathcal{S}$. Therefore, it can be easily shown that
    \begin{itemize}
        \item if the non-zero coefficients of $\mathbf{X}$ are drawn from Gaussian distribution $\mathcal{N}(0,1)$, then $\mathbb{E}[\|\mathbf{X}\|_2^2]=K$,
        \item if the non-zero coefficients of $\mathbf{X}$ are drawn from the uniform distribution $\mathcal{U}[-1,1]$, then $\mathbb{E}[\|\mathbf{X}\|_2^2]=\frac{K}{3}$.
    \end{itemize}

  \item Randomly generate a set of sensing matrix $\mathbf{\Phi}$. We let the elements of the sensing matrix be $\mathbf{\Phi}_{ij} \overset{\textrm{iid}}{\sim} \mathcal{N} (0,1/N)$, and then normalize the columns of $\mathbf{\Phi}$ to unit-norm. Note that once $\mathbf{\Phi}$ is generated, it remains fixed and known to the sparse reconstruction algorithm.
  \item Compute linear measurements $\mathbf{Y= \Phi X}$ for each sparse data, and apply a sparse reconstruction algorithm (here OMP) to acquire the locally reconstructed vector $\bar{\mathbf{x}}$ as discussed in \remref{rem:MMSE estimator}.
  \item Choose the total quantization bit-rate $R_x = Mr_x$ bits/vector where $r_x \in \mathbb{R}^+$ is the allocated rate to a component of $\mathbf{X}$, and design codebook sets for each quantization scheme as described in \remref{rem:codebook}. In order to make a fair comparison and guarantee that the performance of the AbS-based quantization does not decline as compared with that of the nearest-neighbor coding, \algref{alg1} is initialized with the nearest-neighbor codepoints.
  \item Next, apply the quantization algorithms on the generated data $\mathbf{X,Y}$, and assess NMSE by averaging over all data.
\end{enumerate}

\subsection{Experimental Results: Evaluation of the Proposed Schemes} \label{subsec:results} In our
simulations, for implementing \algref{alg1} using the subroutines $\texttt{AbS\_seq}$ and $\texttt{AbS\_nonseq}$, we choose the stopping threshold $\gamma = 10^{-6}$, where we have observed that
\algref{alg1} converges in at most $5$ iterations for all simulation setups. We perform the simulation by averaging over 1000 realizations of $\mathbf{X}$ with non-zero coefficients drawn from the standard Gaussian distribution. Furthermore, the OMP reconstruction algorithm is used to recover the source from quantized measurements.

Next, recall that the number of quantization bits assigned for each entry of $\mathbf{Y} \in \mathbb{R}^N$ is $r_y = \frac{Mr_x}{N}$, and may not be an integer. Hence, in order to utilize all available bits for quantization of CS measurements, we first assign $\lfloor \frac{Mr_x}{N}\rfloor$ bits to all entries, and then allocate another extra bit to the first $Mr_x - \lfloor \frac{Mr_x}{N}\rfloor N$ entries of $\mathbf{Y}$, where $\lfloor \cdot \rfloor$ denotes the floor operator. A similar approach is employed for the direct and support-set coding schemes. Furthermore, using the direct coding (or, support set coding), if $r_x < 1$ bit (or, $r_x < \frac{K}{M} \log_2 M + \frac{K}{M}$ bit), we quantize the first $M r_x$ (or, $K \log_2 M + K$) coefficients of the reconstructed signal $\bar{\mathbf{x}}$ by 1 bit, and the remaining coefficients at the decoder are filled with zero.

\begin{figure}
  \begin{center}
  \includegraphics[width=\columnwidth,height=7.5cm]{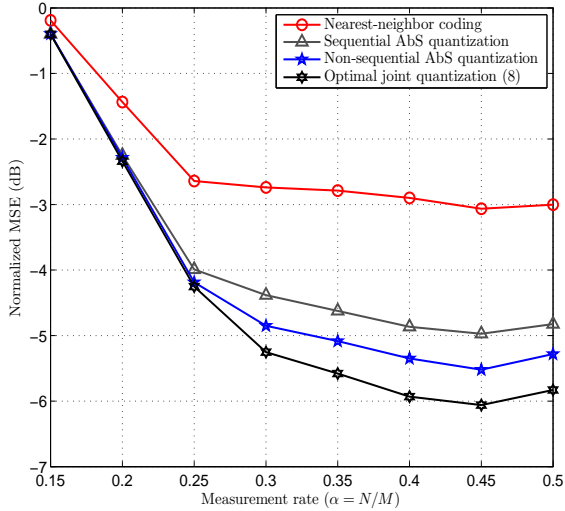}\\
  \caption{Normalized MSE (in dB) as a function of measurement rate ($\alpha = N/M$) given a fixed quantization bit-rate using nearest-neighbor coding, sequential and non-sequential AbS quantization schemes and joint quantization \eqref{eq:final enc original} as a benchmark. The parameters are chosen as $M=40$, $K=3$ and $r_x = 0.5$ bit per component of the source vector.}
  \label{fig:joint quantization}
  \end{center}
\end{figure}

In order to offer insights into the efficiency of our proposed AbS quantization algorithms, we compare their performances with that of the (optimal) joint quantization \eqref{eq:final enc original} which provides a benchmark performance. As mentioned earlier, the complexity of the joint quantization scheme \eqref{eq:final enc original} grows exponentially with the total bit budget. Due to this exponential complexity, a benchmark performance evaluation is impossible to realize at high rate and dimension. However, in order to gain insights regarding the performance of our AbS-based quantizers, we practically implemented the joint optimization method \eqref{eq:final enc original} at very low quantization bit-rate and dimension. The curves in \figref{fig:joint quantization} illustrate the performance (NMSE) as a function of measurement rate $\alpha$ for the simulation setup $M=40$, $K=3$ and $r_x=0.5$ bit per source component. As can be seen, while the nearest-neighbor coding gives a poor performance, the AbS-based quantizers perform nearly optimal, and the gap between the performance of the non-sequential/sequential AbS quantizer and the benchmark joint quantizer is less than 0.5 dB/1 dB in the worst case (at $\alpha=0.45$).

\begin{figure}
  \begin{center}
  \includegraphics[width=\columnwidth,height=7.5cm]{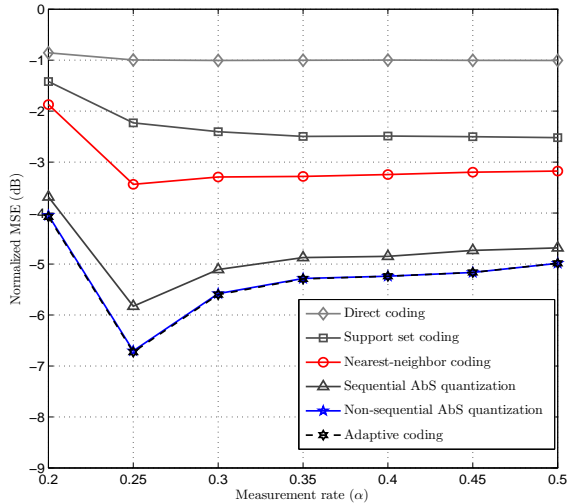}\\
  \caption{Normalized MSE (in dB) as a function of measurement rate ($\alpha = N/M$) given a fixed quantization bit-rate using different quantization schemes. The parameters are chosen as $M=512$, $K=35$ and $r_x = 0.5$ bit per component of the input vector which is equivalent to total bit budget $R_x=256$ bits.}
  \label{fig:result_MSE_fom_new}
  \end{center}
\end{figure}

\begin{figure}
  \begin{center}
  \includegraphics[width=\columnwidth,height=7.5cm]{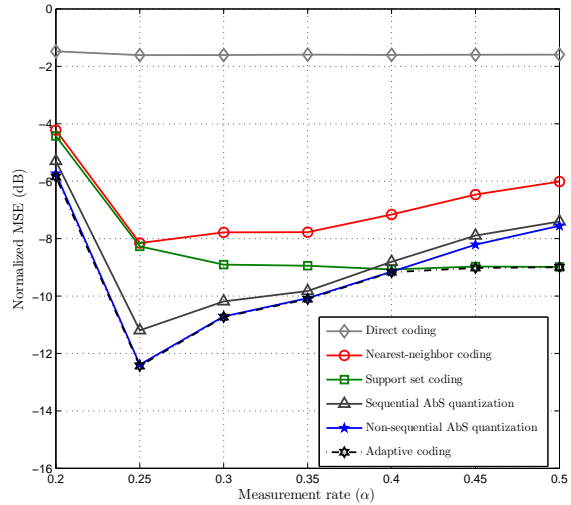}\\
  \caption{Normalized MSE (in dB) as a function of measurement rate ($\alpha = N/M$) given a fixed quantization bit-rate using different quantization schemes. The parameters are chosen as $M=512$, $K=35$ and $r_x = 0.75$ bit per component of the input vector which is equivalent to total bit budget $R_x=384$ bits.}
  \label{fig:result2}
  \end{center}
\end{figure}

Using a larger simulation parameter set $M=512$, $K=35$ (sparsity
ratio $\approx 6.8\%$), we illustrate the performance (NMSE) of the quantization algorithms as a function of measurement rate $\alpha$, shown in
\figref{fig:result_MSE_fom_new} and \figref{fig:result2} for fixed quantization bit-rates $r_x = 0.5$ and $r_x=0.75$ bit per component of $\mathbf{X}$, respectively. First, let us interpret the performance behavior of the schemes for quantization in the CS measurement domain, i.e.,
nearest neighbor coding and the proposed AbS quantization schemes. It
is worth pointing out that increasing the measurement rate $\alpha$,
given a fixed bit budget $r_x$, has two different effects on the
performance. On one hand, it facilitates a more precise reconstruction
both at the encoder and the decoder due to increasing number of
measurements. On the other hand, it reduces (increases) quantization
bit-rate (quantization noise) per entry of the measurement vector since
$r_y = r_x/\alpha$. Following these facts, it can be observed from the
curves that given a very small values of $\alpha$, the sparse reconstruction algorithm
fails to detect the sparsity pattern and reconstruct the source. This results in a poor performance although the quantization bit-rate per entry
is high. As $\alpha$ increases to a certain amount, the reconstruction
algorithm succeeds to reconstruct the sparse source precisely out of
the measurements since the number of measurements is sufficient, and
the quantization noise is small enough. At this point ($\alpha = 0.25$
for \figref{fig:result_MSE_fom_new} and \figref{fig:result2}), the curves reach the best
performance. However, for higher $\alpha$'s, due to the limited
quantization bit-rate $r_y$, the quantization noise per entry increases
which leads to a poorer performance. Among these schemes, the
performance of the nearest-neighbor coding is the worst since it
does not take the end-to-end distortion into account. However, the
sequential and non-sequential AbS-based algorithms improve the
performance significantly by exploiting the AbS framework. The gap
between the performance of the two AbS schemes reduces as $\alpha$
increases since CS measurements tend to become i.i.d. RV's, and therefore, the non-sequential optimization method does not provide any extra gain. Observing the curves in \figref{fig:result_MSE_fom_new} and \figref{fig:result2}, it can be also found that the MSE-minimizing measurement rate using both $r_x = 0.5$ and $r_x = 0.75$ occurs at $\alpha = 0.25$. We cannot, in general, claim that the optimal $\alpha$ would be the same at all quantization bit-rate regions. However, it can be inferred that the curves reach their minima, and then they take an upward trend.

Next, we evaluate the performance of the schemes for quantization in
the reconstructed signal domain. At small $\alpha$'s, the
reconstruction algorithm fails to reconstruct the locally sparse
source. It can be seen that as $\alpha$ increases, and the sparse reconstruction
algorithm is able to reconstruct the input signal vector, the curve
does not vary much by further increasing measurement rate since the
allocated quantization bits using this scheme are independent of
number of measurements, unlike the quantization schemes for
measurement domain. The direct coding leads to a poor performance
since this scheme does not take into consideration the sparsity
pattern, while the support set coding improves the performance by only
quantizing the magnitude of the non-zero coefficients in the estimated
support set. By adaptively choosing the better performance among the
quantization schemes in signal and CS measurement domains, one can
benefit from both schemes at all ranges of $\alpha$ which is labeled
by adaptive coding. However, note that in the spirit of exploiting CS
for practical applications, we are mainly interested in the lower
ranges of $\alpha$, for example, at $\alpha = 0.25$ in
\figref{fig:result2}, where the AbS quantization schemes achieve at
least a considerable $3$ dB reduction in MSE compared to other
competing schemes.

In addition to the experimental results presented in \figref{fig:result_MSE_fom_new} and \figref{fig:result2}, we also carried out similar simulations with the parameter set $M=1000$, $K=40$ (sparsity ratio $=4\%$), $r_x = 0.4$ bit per component of $\mathbf{X}$. We found similar trend in performances of all competing schemes and that the AbS quantization algorithms improve the performance at least 3 dB compared to the nearest-neighbor coding. Nevertheless, the direct and support set coding schemes achieve poor performances.

\begin{figure*}
  \begin{center}
  \includegraphics[width=0.9\linewidth,height=12cm]{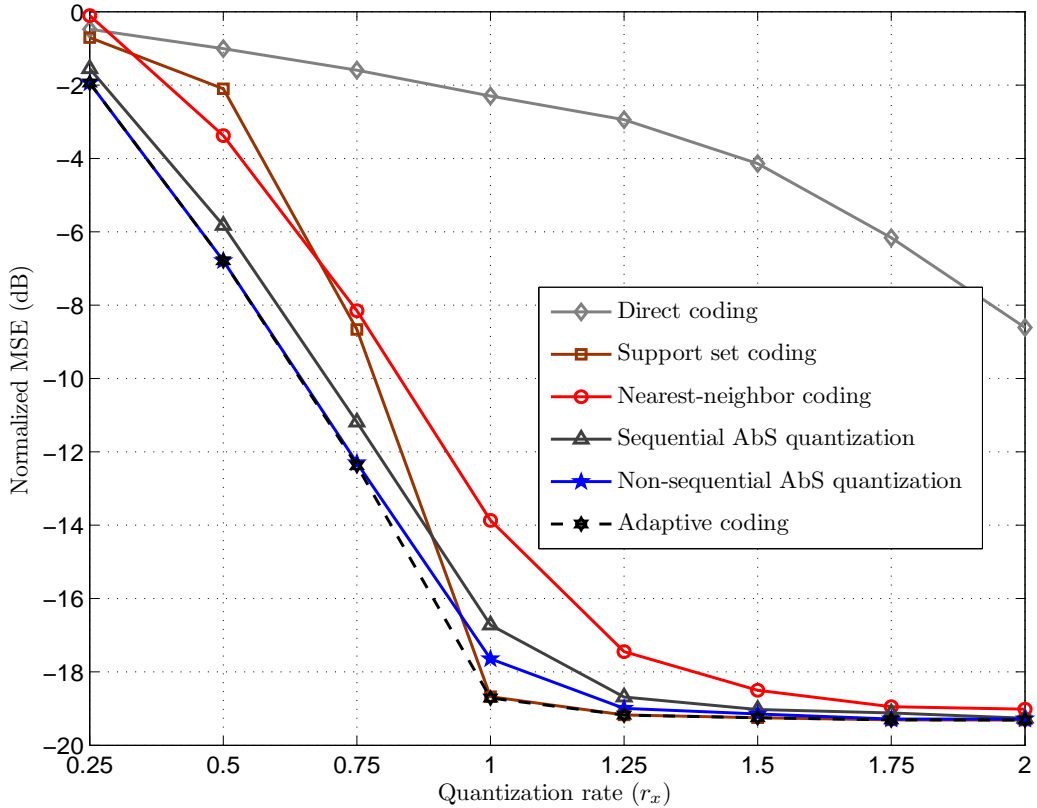}\\
  \caption{Normalized MSE (in dB) as a function of rate per each component of $\mathbf{X}$ (i.e., $r_x$) given a fixed measurement rate using different quantization schemes. The parameters are chosen as $M=512$, $K=35$ and $\alpha=0.25$. The values on the x-axis are equivalent to $r_y=1$ bit to $r_y=8$ bits per measurement entry. At very high rates, the curves achieves MSE floor due to the fixed sparse reconstruction distortion.}
  \label{fig:result4}
  \end{center}
\end{figure*}

Next, we examine the performance (NMSE) as a
function of quantization bit-rate per entry of $\mathbf{X}$, i.e. $r_x$,
which is reported for the simulation setup $M=512$, $K=35$ at fixed
$\alpha=0.25$ in \figref{fig:result4}. It can be observed that in low
and moderate bit-rate regimes, the AbS quantization outperforms the other
schemes, while at high bit-rates the support set coding attains a slightly
better performance. Using the adaptive coding scheme, one can achieve
the best performance among the competing schemes at all ranges of
quantization bit-rate. At very high bit-rates, it is observed that the NMSE curves saturate. This is due to the fact that although the distortion due to quantization decreases as quantization bit-rate increases, the distortion due to sparse reconstruction (because of low number of measurements) still exists at a fixed measurement rate. At very high bit-rates, in order to eliminate the MSE floor, we need more number of measurements (sensors) so that the sparse reconstruction distortion becomes negligible. However, note that the use of more number measurements is restricted in CS.

\subsection{Experimental Results: Comparison with Existing Schemes}
In order to verify the efficiency of our proposed AbS schemes, we compare the performance of the sequential AbS quantizer vis-a-vis LASSO-optimized quantizer and uniform quantizer of \cite{09:Sun}. It is note-worthy while our design method is neither asymptotic nor limited to any particular sparse reconstruction algorithm, the LASSO-optimized quantizer of \cite{09:Sun} is based on two main assumptions: (1) asymptotic quantization bit-rate, (2) LASSO reconstruction for recovering a sparse source from quantized measurements. Another important design difference between the LASSO-optimized quantizer and our AbS-based quantizer schemes is described as follows. The LASSO-optimized scheme is based on the state-of-the-art distributed functional scalar quantizer (DFSQ) \cite{08:Vinith} for quantization of a scalar function. Therefore, it only minimizes the MSE of a scalar function (arbitrary output of the LASSO sparse reconstruction) of the measurement vector. However, in our AbS schemes, we design the quantizers by taking into consideration the MSE between a source vector and its final reconstruction vector.

In order to reconstruct a sparse source $\widehat{\mathbf{x}}$ from quantized measurements $\widehat{\mathbf{y}}$, the LASSO reconstruction algorithm aims at solving the following convex optimization problem
\begin{equation} \label{eq:LASSO}
    \widehat{\mathbf{x}}_{\text{LASSO}}= \underset{\mathbf{x} \in \mathbb{R}^M}{\textrm{argmin}} \hspace{0.1cm} \|\widehat{\mathbf{y}} - \mathbf{\Phi x}\|_2  + \mu \|\mathbf{x}\|_1,
\end{equation}
where $\mu > 0$ is a fixed user parameter. We solve \eqref{eq:LASSO} using the SPGL1 toolbox \cite{spgl1:2007}.

\begin{figure}
  \begin{center}
  \includegraphics[width=\columnwidth,height=7.5cm]{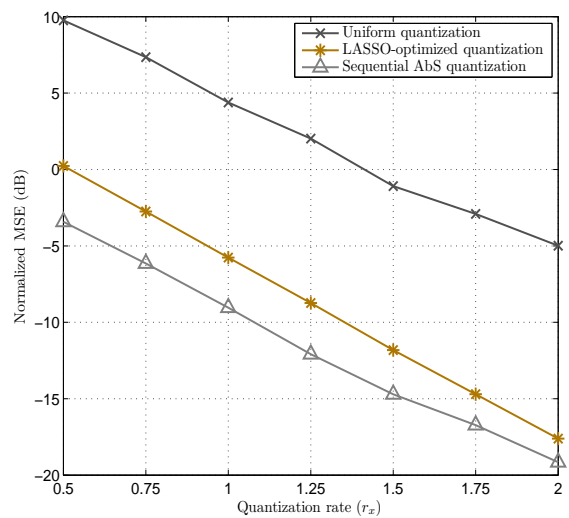}\\
  \caption{Normalized MSE (in dB) as a function of rate per each component of $\mathbf{X}$ (i.e., $r_x$) given a fixed measurement rate $\alpha=0.5$ using the proposed sequential AbS, LASSO-optimized and uniform quantizer schemes of \cite{09:Sun}. The parameters are chosen as $M=100$, $K=10$ and $\alpha=0.5$.}
  \label{fig:comparison LASSO}
  \end{center}
\end{figure}

For fair comparison, we use the same kind of sources as used in the schemes of \cite{09:Sun}. We randomly generate the non-zero coefficients of $\mathbf{X}$ from uniform distribution $\mathcal{U}[-1,1]$, and use the LASSO reconstruction \eqref{eq:LASSO} with $\mu = 10^{-3}$ (the choice of $\mu$ is experimentally verified to achieve the best performance). All other simulation setups are as the same as those given in \secref{subsec:setup}. Also, simulation parameters are set to $M=100$, $K=10$ and $\alpha=0.5$. In \figref{fig:comparison LASSO}, the performance (in terms of NMSE) as a function of quantization bit-rate $r_x$ is illustrated for the sequential AbS quantizer as well as LASSO-optimized quantizer and uniform quantizer design schemes. As would be expected, the uniform quantizer provides a poor performance because of inappropriate codepoints for quantization of measurement entries. It can be also seen that the proposed sequential AbS quantization gives a better performance (almost 3 dB) than the LASSO-optimized quantizer.

\section{Conclusions} \label{sec:conclusion}
Due to non-linear behavior of sparse reconstructions, the effect of quantization on CS measurements would reflect in a non-linear manner in the signal reconstruction. To handle the non-linearity, we have developed AbS-based quantization schemes, and shown that a significant improvement in performance can be achieved. We have found that the AbS schemes have a limitation at high quantization bit-rates, and hence we have developed an adaptive coding scheme suited for all scenarios. Furthermore, comparisons with existing quantization algorithms for CS measurements have demonstrated the efficiency of our AbS-based quantization schemes.

\appendix

We briefly describe the orthogonal matching pursuit (OMP) algorithm \cite{07:Tropp,08:Blumensath} for reconstructing a sparse source from quantized CS measurement vector $\widehat{\mathbf{y}}$, where the sensing matrix and sparsity level are provided as well. The low-complexity OMP scheme is an iterative algorithm where it performs a matched filter operation and an orthogonal projection at each iteration. Using pseudo-inverse matrix inversion, the orthogonal projection operations can be performed recursively.  The main steps of the OMP are summarized in~\algref{alg:OMP}.
\begin{algorithm}[ht!]
\caption{: Orthogonal matching Pursuit (OMP)}\label{alg:OMP}
\begin{algorithmic}[1]
\STATE \textbf{input: } $\mathbf{\Phi}$, $\mathbf{\widehat{\mathbf{Y}} = \widehat{\mathbf{y}}}$, $K$
\STATE \textbf{initialization: }$l \leftarrow 0$ (Iteration counter )
$\mathbf{r}^{(0)} \leftarrow \widehat{\mathbf{y}}$ (Initial residual), $\mathcal{S}^{(0)} \leftarrow \varnothing$ (Initial support set)
\REPEAT
\STATE $l \leftarrow l+1$
\STATE $i^{(l)} \leftarrow $ index of the highest amplitude of $\mathbf{\Phi}^{T} \mathbf{r}^{(l-1)}$ \label{step:atom_index_OMP}
\STATE $\mathcal{S}^{(l)} \leftarrow \mathcal{S}^{(l-1)} \cup i^{(l)}$ \hfill (Note: $|\mathcal{S}^{(l)}|=l$) \label{step:Intermediate_Support_OMP}
\STATE $\mathbf{r}^{(l)} \leftarrow \widehat{\mathbf{y}}-\mathbf{\Phi}_{\mathcal{S}^{(l)}} \mathbf{\Phi}_{\mathcal{S}^{(l)}}^{\dag} \widehat{\mathbf{y}}$ \hfill (Orthogonal projection)	 \label{step:Proj_Residue_OMP}
\UNTIL $( (\| \mathbf{r}^{(l)} \|_{2} > \| \mathbf{r}^{(l-1)} \|_{2}) \,\, \mathrm{or}\,\, (l>K) )$
\STATE $l \leftarrow l-1$ \hfill (Previous iteration)
\STATE \textbf{output:} $\widehat{\mathbf{x}} \! \in \! \mathbb{R}^{M}$, satisfying $\widehat{\mathbf{x}}_{\mathcal{S}^{(l)}} \!=\! \mathbf{\Phi}_{\mathcal{S}^{(l)}}^{\dag} \widehat{\mathbf{y}}$ and $\widehat{\mathbf{x}}_{\overline{\mathcal{S}}^{(l)}} \!=\! \mathbf{0}$
\end{algorithmic}
\end{algorithm}

\bibliographystyle{IEEEtran}
\bibliography{IEEEfull,bibliokthPasha}

\end{document}